\documentclass[useAMS,usenatbib]{mn2e}

\title[The 2SLAQ survey: Evolution of LF to $z=0.6$]{The 2dF-SDSS LRG and QSO survey: Evolution of the Luminosity Function of Luminous Red Galaxies to $z=0.6$}
\author[Wake et al.]{ \parbox{\textwidth}{
David A. Wake$^{1,2}$,
Robert C. Nichol$^{1}$,
Daniel J. Eisenstein$^{3}$,
Jon Loveday$^{4}$,
Alastair C. Edge$^2$,
Russell Cannon$^5$,
Ian Smail$^{6}$,
Donald P. Schneider$^7$,
Ryan Scranton$^{15}$,
Daniel Carson$^1$,
Nicholas P. Ross$^2$, 
Robert J. Brunner$^8$,
Matthew Colless$^5$,
Warrwick J. Couch$^{9}$,
Scott M. Croom$^5$,
Simon P. Driver$^{10}$,
Jos\'{e} da \^{A}ngela$^2$,
Sebastian Jester$^{14}$,
Roberto de Propris$^{11}$,
Michael J. Drinkwater$^{12}$,
Joss Bland-Hawthorn$^5$,
Kevin A. Pimbblet$^{12}$,
Isaac G. Roseboom$^{12}$,
Tom Shanks$^2$,
Robert G. Sharp$^5$,
Jon Brinkmann$^{13}$
}
  \vspace*{6pt} \\
$^{1}$Institute of Cosmology and Gravitation (ICG), Mercantile House, University of Portsmouth, Portsmouth, PO1 2EG, UK\\
$^{2}$Dept. of Physics, Durham University, South Road, Durham, DH1 3LE, UK\\
$^{3}$Steward Observatory, University of Arizona, 933 N. Cherry Ave., Tucson 
AZ85721, USA\\
$^{4}$Astronomy Center, University of Sussex, Falmer, Brighton, BN1 9QH, UK\\
$^5$Anglo--Australian Observatory, PO Box 296, NSW 1710, Australia\\
$^6$Institute of Computational Cosmology, Durham University, South Road, Durham, DH1 3LE, UK\\
$^7$Department of Astronomy and Astrophysics, Pennsylvania State University, 525 Davey Laboratory, University Park, PA 16802, USA\\
$^8$National Center for Supercomputing Applications, University of Illinois at Urbana-Champaign, 1205 W. Clark St., Urbana, IL 61801, USA\\
$^{9}$School of Physics, University of New South Wales, Sydney, Australia\\
$^{10}$School of Physics \& Astronomy, University of St Andrews, North Haugh, St Andrews, KY16 9SS, UK\\
$^{11}$Cerro Tololo Inter-American Observatory, Casilla 603, La Serena, Chile\\
$^{12}$Dept. of Physics, University of Queensland, Brisbane, QLD 4072, Australia\\
$^{13}$Apache Point Observatory, 2001 Apache Point Road, P.O. Box 59, Sunspot, NM 88349, USA\\
$^{14}$ School of Physics \& Astronomy, University of Southampton Southampton, SO17 1BJ, UK\\
$^{15}$Dept. of Physics and Astronomy, University of Pittsburgh,  3941 O'Hara Street, Pittsburgh, PA 15260, USA\\}

\begin{document}

\pagerange{\pageref{firstpage}--\pageref{lastpage}} \pubyear{2002}

\maketitle

\label{firstpage}

\begin{abstract}
  We present new measurements of the luminosity function (LF) of
  Luminous Red Galaxies (LRGs) from the Sloan Digital Sky Survey
  (SDSS) and the 2dF--SDSS LRG and Quasar (2SLAQ) survey. We have
  carefully quantified, and corrected for, uncertainties in the K and evolutionary
  corrections, differences in the colour selection methods, and the
  effects of photometric errors, thus ensuring we are studying the
  same galaxy population in both surveys. Using a limited subset of
  6326 SDSS LRGs (with $0.17<z<0.24$) and 1725 2SLAQ LRGs (with $0.5
  <z<0.6$), for which the matching colour selection is most reliable,
  we find no evidence for any additional evolution in the LRG LF, over
  this redshift range, beyond that expected from a simple passive
  evolution model. This lack of additional evolution is quantified
  using the comoving luminosity density of SDSS and 2SLAQ LRGs,
  brighter than $M_{^{0.2}r}-5logh_{0.7}=-22.5$, which are 
  $2.51\pm0.03\,\times
  10^{-7}\,{\rm L_{\odot}\, Mpc^{-3}}$ and $2.44\pm0.15\,\times
  10^{-7}\,{\rm L_{\odot}\, Mpc^{-3}}$ respectively ($<10\%$
  uncertainty). We compare our LFs to the COMBO-17 data and find
  excellent agreement over the same
  redshift range.  Together, these surveys show no evidence for
  additional evolution (beyond passive) in the LF of LRGs brighter
  than M$_{^{0.2}r}-5logh_{0.7}=-21$ (or brighter than $\sim L^{\star}$).  We
  test our SDSS and 2SLAQ LFs against a simple ``dry merger'' model
  for the evolution of massive red galaxies and find that at least half 
  of the LRGs at $z\simeq0.2$ must already have been  well-assembled 
  (with more than half their stellar mass) by $z\simeq0.6$. 
  This limit is barely consistent with recent results from semi-analytical 
  models of galaxy evolution.
\end{abstract}


\section{Introduction}

The formation of massive elliptical galaxies is a major conundrum in
modern cosmology. Observationally, it is clear that such galaxies
formed the bulk of their stars at redshifts greater than two, with
evidence coming from a variety of studies, including observations of
clusters and groups of galaxies
\citep[]{1992MNRAS.254..601B,1997ApJ...483..582E,
  1998A&A...334...99K,1999AJ....118..719D,2002MNRAS.329L..53B,2005ApJ...627..186W,2005ApJ...620L..83H,2006MNRAS.366..645P},
optical/NIR galaxy count and colour studies \citep{met96,met01,met05} and 
spectroscopic surveys of field galaxies over a range
of redshifts
\citep{1998ApJ...508L.143B,2000AJ....120..165T,2003AJ....125.1817B,2003AJ....125.1849B,2003AJ....125.1866B,2003AJ....125.1882B,2002AJ....124..646H,1998MNRAS.295L..29K,2004ApJ...600..681B,2004Natur.430..181G,2004ApJ...614L...9M,2004Natur.430..184C,2005ApJ...621..673T,2005ApJ...631..101P,2006AJ....131.1288B}.

These studies also suggest that the evolution of a majority of these
massive ellipticals is consistent with a simple passive model of
stellar evolution. For example, \citet{2003AJ....125.1866B,
  2003AJ....125.1882B} have used the Sloan Digital Sky Survey 
\citep[SDSS;][]{2000AJ....120.1579Y} to study the evolution of 
luminous early--type
galaxies (out to $z\simeq0.3$) and find that, for a fixed velocity
dispersion, older ellipticals are redder and fainter in luminosity,
fully consistent with the expected fading of their stellar populations
with time.  They also find that the environment (as measured by local
density) of luminous ellipticals has only a weak effect upon their
properties, e.g., the Fundamental Plane fades by only 0.075
mags/arcsec$^2$ from the field to the cores of clusters (consistent
with cluster galaxies forming at a slightly earlier epoch than field
ellipticals). Several other studies have also found very similar results 
\citep{2002MNRAS.337..172K,2006astro.ph..3714C,2006AJ....131.1288B}.  
However, several authors have found evidence for recent
and/or on-going star-formation in a fraction of local massive
ellipticals \citep{2000AJ....120..165T,2003PASJ...55..771G,
  2004ApJ...601L.127F,2005MNRAS.360..587B}. This fraction appears
to increase with redshift \citep{LeBorgne2005,2006Roseboom} and decrease 
with mass \citep{2003AJ....125.2891C,2005ApJ...632..137N,2006astro.ph..3714C}.

One potential problem for studies of massive elliptical/early-type
galaxies, particularly those tracking the evolution with redshift, is
that of {\it progenitor bias}
\citep{1996MNRAS.281..985V,2000ApJ...541...95V}. By studying only
galaxies that look like massive early-type systems at high redshift,
some of the progenitors of the low redshift massive early-type
galaxies may be missed. An effective way to counter this problem is to
study the evolution in the number density, or luminosity function
(LF), of massive early-type galaxies with redshift.  If these galaxies
did indeed form at high redshift, then there will be little evidence
for the evolution of their LF with redshift. Such studies have only
just begun, e.g., the COMBO--17 (C17) survey, which reported on the
evolution of $\simeq 5000$ red (and thus implied early-type) galaxies
and found at most a factor of two evolution out to $z\simeq1$
\citep[corresponding to 9 Gyrs in look--back
time;][]{2004ApJ...608..752B}. This is consistent with earlier (but
smaller in number) studies of high redshift red (early--type) galaxies
from the CFRS \citep{1995ApJ...455..108L,1999ApJ...525...31S}, CNOC2
\citep{1999ApJ...518..533L} and K20 \citep{2003A&A...402..837P}
surveys. At higher redshifts ($z>1$), recent spectroscopic surveys
suggest that a significant fraction of massive galaxies are already
in place at these early epochs
\citep{2004Natur.430..181G,2004Natur.430..184C}.

Theoretically, the existence of massive, passively--evolved
ellipticals has been a major challenge for some models of galaxy evolution.
For example, in the favoured hierarchical Cold Dark Matter (CDM) model
of structure formation, such massive galaxies are expected to reside
in the largest dark matter haloes, that form at late times through the
merger of smaller mass haloes (see \citet{1996MNRAS.281..487K} for
early predictions). In recent years, there have been a number of
prescriptions proposed to solve the ``anti--hierarchical'' nature of
the formation and evolution of massive ellipticals and thus better
match the observations discussed above, including: {\it i)} The
reduction of the gas cooling rate \citep{2003ApJ...599...38B}; {\it
  ii)} Super-winds that eject the gas once it has cooled, but before
it can form stars \citep{2003ApJ...599...38B,2005MNRAS.356.1191B};
{\it iii)} Shock heating of infalling gas and -PdV work of the gas
\citep{2005astroph0512235}; {\it iv)} Feedback
from Active Galactic Nuclei
\citep[AGN;][]{2005MNRAS.358L..16K,2005apj...635L..13S}.

The latter of these proposed effects (AGN feedback) is appealing
because of the observed empirical relationship between the
luminosity, central concentration and mass of galactic bulges and 
the mass of their central super-massive black holes 
\citep{2000ApJ...539L...9F,2002ApJ...574..740T,2006ApJ...637...96N}.
 AGN feedback comes in
two flavors: {\it i)} The merger of gas--rich galaxies causing an
initial starburst, followed by the growth of the central black hole
and a ``quasar wind'' which quenches further star--formation
\citep{hopkins2005}; {\it ii)} Radio feedback from low luminosity
AGN that suppresses the cooling of gas in massive halos resulting in the
termination of late--time star-formation
\citep{2005astroph0511338,2006MNRAS.365...11C}.  Detailed simulations
of these AGN feedback models on the evolution of ellipticals has
helped resolve the discrepancy between present observations and the
naive hierarchical expectations of the CDM model
\citep{2005ApJ...620L..79S,2005astroph0509725,2005astroph0511338,hopkins2005,2006MNRAS.365...11C}.
For example, \citet{2005astroph0509725} predict that for galaxies
more massive than $10^{11}{\rm M_{\odot}}$, over 50\% of their stars
have formed by a median redshift of $z=2.6$, while the typical
assembly redshift of these galaxies (when these stars reside in a
single object) is only $z\simeq 0.8$. This requires ``dry mergers'' of
the smaller galaxies (i.e. without gas) to avoid causing bursts of new
star--formation. Several authors have found evidence for such dry mergers
\citep[see][]{2005AJ....130.2647V, 2005ApJ...625...23B}.

In this paper, we present an accurate measurement of the evolution of
the luminosity function (LF) of Luminous Red Galaxies
\citep[LRGs;][]{2001AJ....122.2267E}, that are predominantly massive
elliptical galaxies and thus provide strong constraints on the
physical mechanisms that have been suggested to produce
anti-hierarchical galaxy formation in a CDM universe. We utilise two
samples of LRGs, one from the SDSS ($0.15 < z < 0.37$), and a new
survey of higher redshift LRGs ($0.45 < z < 0.8$) selected from the
multi--colour SDSS photometry, but spectroscopically observed using
the 2dF spectrograph on the Anglo--Australian Telescope (AAT). This
new survey is known as the 2dF-SDSS LRG and QSO (2SLAQ) survey.

The LRG component of the 2SLAQ survey is a significant advance over
previous surveys of massive early-type galaxies due to an increase in
both the volume surveyed ($\sim10^7\,h^{-3}\,{\rm Mpc^{3}}$), thus
reducing the problem of cosmic variance, and the number of galaxy
redshifts obtained, thus greatly increasing the statistical accuracy of
the LF at bright magnitudes.  In Section \ref{sec:data}, we describe
the 2SLAQ survey, while in Section 3 we present a detailed discussion
of the K and evolutionary corrections, photometric errors and consistent colour
selections for the SDSS and 2SLAQ samples. In Section \ref{sec:lf} we
present the luminosity functions of the SDSS and 2SLAQ surveys. In
Section \ref{sec:dis}, we discuss our results in terms of models of
galaxy evolution and conclude. Throughout, we adopt a cosmology of
$\Omega_M$ = 0.3, $\Omega_{\Lambda}$ = 0.7 and $H_0$ = 70 km s$^{-1}$
Mpc$^{-1}$.

\section{LRGs in the SDSS and 2SLAQ surveys}
\label{sec:data}

We present in this paper an analysis of LRGs taken from both the
SDSS survey and the 2dF--SDSS LRG and QSO (2SLAQ) survey. For the SDSS
data, we only use the LRG Cut I photometric selection (using the $g-r$
and $r-i$ colours of galaxies) as defined and discussed in Eisenstein
et al. (2001; E01). This provides us with a pseudo volume--limited sample
of LRGs, with $M_r \le -21.8$ and $0.15 < z < 0.35$, selected from the
SDSS Data Release \citep[DR3,][]{2005astroph0507711} dataset.  Below
$z = 0.15$, the space density of SDSS LRGs increases by nearly 50\%
because of contamination by low redshift star--forming galaxies in the
colour selection, while above $z>0.35$, the space density of SDSS LRGs
begins to decrease because of the flux limit and the degeneracy
between the colours and redshift of LRGs close to $z\simeq0.37$ where
the $4000$\AA break feature passes out of the SDSS $g$--band into the
$r$--band \citep{1996AJ....111.1748F}.  Within the range
$0.15<z<0.35$, the space density of SDSS Cut I LRGs is approximately
constant with redshift (see E01).

In 2003, the 2SLAQ survey began with the goal of producing a pseudo
volume--limited sample of 10,000 LRGs, with a median redshift of
$z=0.55$, and 10,000 faint $z<3$ quasars, both selected from the SDSS
multi--colour imaging data. In this paper, we focus on the 2SLAQ LRGs
which are selected using similar criteria as the Cut II SDSS LRGs in
E01. The key differences are: {\it (i)} The apparent magnitude limit
has been lowered to $m_i{\rm (model)} < 19.8$, thus extending the
volume--limited LRG samples to $z\simeq0.6$; {\it (ii)} The effective
rest--frame colour cut ($c_{\perp}$ in E01) has been shifted slightly bluer
than in the SDSS LRG selection to accommodate the density of 2dF
fibres available.  This provides a less conservative colour cut, at
these higher redshifts, which is essential for studying potentially
small changes in their colour.

The details of the 2SLAQ LRG selection and observations are presented
elsewhere \citep{2006Cannon}. However, we have measured over 11000 LRG
redshifts, covering $180 {\rm deg^2}$ of SDSS imaging data, from 87
allocated nights of AAT time.  Over 90\% of these galaxies are within
the range $0.45<z<0.7$. The targeted LRGs where split into three
subsamples as detailed in \citet{2006Cannon}, with the primary sample
(Sample 8) accounting for two thirds of these. We only focus on Sample
8 in this paper due to its high completeness and uniform selection.
The overall success rate of obtaining redshifts from the 2dF spectra
for Sample 8 LRGs is 95\%, while the centers of the 2dF fields were
spaced by 1.2$^{\circ}$, resulting in an overall redshift completeness
of sample 8 LRG targets of $\sim$75\% across the whole survey area
(see section \ref{sec:comp}).  These data have recently been used to
calibrate LRG photometric redshifts
\citep{2005MNRAS.359..237P,2006collister}.

Although the SDSS magnitude system was designed to be on the AB scale
\citep{ 1996AJ....111.1748F}, the final calibration has differences
from the proposed values by a few percent. We have applied the
corrections m$_{AB}$ = m$_{SDSS} + [-0.036, 0.012, 0.010, 0.028,
0.040]$ for $u$, $g$, $r$, $i$, $z$ respectively (Eisenstein, priv.
comm.).  All magnitudes and colours presented throughout this paper
are corrected for Galactic extinction \citep{1998ApJ...500..525S}.

\section{Sample Selection}

The galaxy samples discussed in this paper were originally selected
under the assumption that LRGs are old, passively--evolving galaxies.
Here, we continue this fundamental assumption when investigating the
evolution of the LRG luminosity function by applying the same
passively--evolving models to our data when computing K and
evolutionary (e) corrections, luminosity functions and colour
selection boundaries. If this assumption is correct, then our K+e
corrected luminosity functions will be identical over the whole
redshift range studied here, with any additional evolution in the LRG
population (beyond the simple passive model) being displayed as a
change in the LF with redshift.  However, as we are using two slightly
different selections of LRGs to generate our total LRG sample, it is
vital that we select the same galaxy population at all redshifts.

\subsection{K+e corrections}
\label{sec:KEcorr}

In order to make a fair comparison between the SDSS and 2SLAQ LRG
samples (at different redshifts), we must correct the properties of
our observed galaxies (magnitude, colours {\it etc.}) into rest--frame
quantities by applying K--corrections. In addition, we can also
correct these rest--frame quantities for the expected evolutionary
changes over the redshift range studied. These e--corrections are
usually performed assuming a model for the galaxy spectral energy
distribution (SED) and its evolution with redshift. In this paper, we
therefore generate our K+e corrections using the
\citet{2003MNRAS.344.1000B} stellar population synthesis code. In
detail, we generate two stellar population models the first of which
forms all its stars in a single instantaneous burst at $z$ = 9.84
(solar metallicity) and then evolves passively with no further star
formation. The second model forms the bulk of its stars in a similar
burst, but includes a small amount of continuous star formation
throughout the rest of its evolution, accounting for 5\% of its final
mass. These two models are shown in Figure 1 and are labelled
``Passive'' and ``Passive+SF'' respectively.

\begin{figure}
\vspace{12.0cm}
\includegraphics{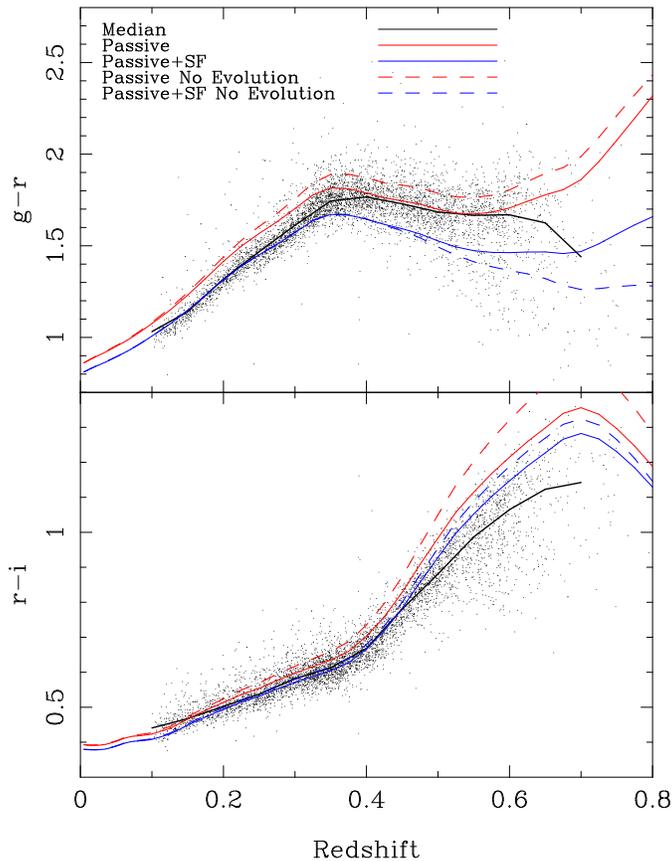}
\caption{\label{fig:colz} The $g-r$ and $r-i$ colours of the SDSS and 2SLAQ LRGs as a function of redshift (points). The solid black line shows the median of the fits with the coloured lines showing the model tracks described in the text with ({\it solid}) and without ({\it dashed}) evolution applied.}
\end{figure}
  
Figure \ref{fig:colz} shows the colour evolution as a function of
redshift for these two models along with the actual measured colours
of the SDSS and 2SLAQ LRGs. Some differences between the models and
data are evident, e.g., the models are too red in $g-r$ and too blue
in $r-i$ for the lowest redshift LRGs, with the opposite effect for
the highest redshift LRGs. This offset between models and data in the
lower redshift LRGs was noted in E01 and a correction to the $g-r$
colours of 0.08 magnitudes was applied. However, with the addition of
the higher redshift sample, it is clear that a simple offset is an
inadequate correction over our entire redshift range. Such differences
between the models and observed colours of early-type galaxies has
been seen before, e.g., \citet{2005ApJ...627..186W} were unable to
match the red sequence colours in galaxy clusters at similar
redshifts, while a similar offset is seen in the colour--redshift
plots of \citet{2005ApJ...635..243F} from the GOODS/CDFS fields.
Simple changes to the models, such as the formation redshift,
metallicity and IMF of the models, are unable to improve the model
fits to the data.  We also note that using the PEGASE stellar
population synthesis model \citep{1997A&A...326..950F} produced very
similar results. It appears that the models are not accurately
reproducing the shape of the spectrum between $4000$\AA\ and
$5000$\AA\, causing an offset in $g-r$ and $r-i$ as the $r$ filter
passes through this (rest--frame) wavelength region. The $g-i$ colours
of LRGs are well reproduced by the models.

\begin{figure}
\vspace{6.5cm}
\includegraphics{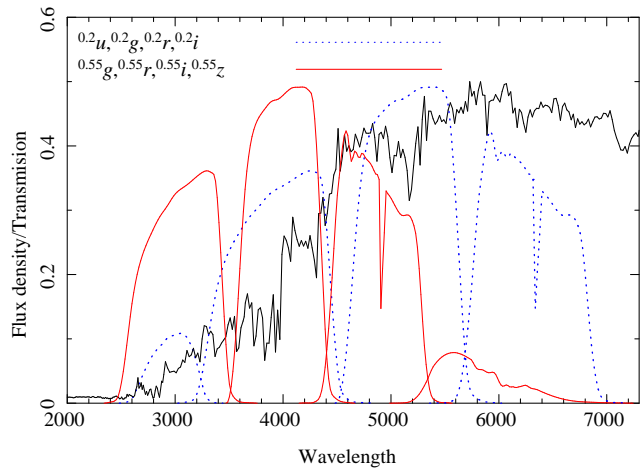}
\caption{\label{fig:filz} The spectral energy distribution of the passive LRG model described in the text, along with the filter transmission curves blueshifted to $z$ = 0.2 and $z$ = 0.55.}
\end{figure}

In order to minimise the systematic uncertainties in the models, and
thus uncertainties in our K+e corrections, we will exploit the fact
that the redshifted SDSS $g$, $r$, $i$ and $z$ passbands at $z=0.55$
(i.e., the median redshift of the 2SLAQ LRGs) approximately overlap
the $u$, $g$, $r$ and $i$ passbands at $z=0.2$ (the typical redshift
of a SDSS LRG). This effect is illustrated in Figure \ref{fig:filz}
where the transmission curves of the SDSS filters are plotted at $z =
0.2$ and $z = 0.55$ and compared to the rest--frame spectral energy
distribution of our passively evolved stellar population model as
discussed above.

Therefore, for each LRG, we estimate its K+e correction by interpolating
between the two SED models using the observed $g-i$ colour of
the galaxy. For the SDSS LRG sample, we simply correct the observed
$g$, $r$ and $i$ magnitudes to $g$, $r$ and $i$ at $z = 0.2$. The
median size of these K+e corrections ranges from 0.01 in $i$ passband
to 0.43 in $g$--band. For the 2SLAQ LRG sample, we again use the
observed $g-i$ colour to interpolate between the two SED models, but
this time correct from observed $r$, $i$ and $z$ passbands to $g$, $r$
and $i$ at $z = 0.2$. The median size of these K and K+e corrections are
0.32 and 0.7 respectively when going from the $r$--band to the $g$--band, 0.1 and 0.4 
going $i$--band to $r$--band, and 0.1 and 0.4  when correcting from
$z$--band to $i$--band.  Tables of these corrections for the two
different SED models are given in Appendix \ref{sec:K+e}.

Throughout the rest of this paper, we use the notation M$_{^{0.2}r}$
to represent the absolute magnitude of an LRG observed through a SDSS
$r$--band filter redshifted to $0.2$. This notation is similar to that
used in \citet{2003ApJ...592..819B}, except here we include an
evolutionary correction in addition to the K-correction. We note that
M$_{^{0.2}r}$ $\simeq$ M$_r(z=0)$ + 0.11 for the colour of a typical
LRG.

\subsection{Consistent Colour Selection}
\label{sec:sample}

\begin{figure}
  \vspace{16.3cm} \includegraphics{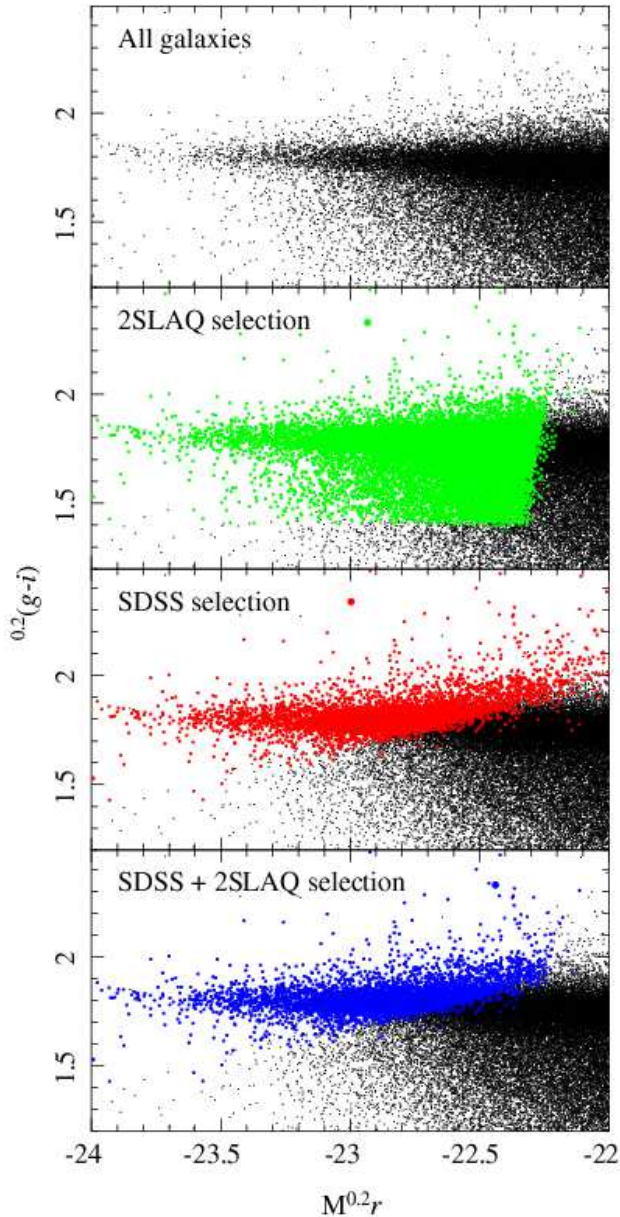}
\caption{\label{fig:selectioncmp}The $^{0.2}(g-i)$ versus M$_{^{0.2}r}$ colour magnitude relation for SDSS main galaxies with $0.1 < z < 0.2$ all K+e corrected to $z$ = 0.2. The small black points in each panel show the whole sample. The second panel shows those galaxies that would have been selected by the 2SLAQ selection criteria when K+e corrected to $z$ = 0.55 (large points). The third panel shows those galaxies that would be selected by the SDSS LRG Cut I selection criteria when K+e corrected to $z$ = 0.2 (large points). The final panel shows the galaxies that satisfy both the SDSS and 2SLAQ LRG selection criteria at both $z$ = 0.2 and 0.55 respectively (large points).  }
\end{figure}

The differences in the SDSS and 2SLAQ selection criteria discussed
above are explicitly illustrated in Figure \ref{fig:selectioncmp}.
Here, we have taken a sample of SDSS main galaxies
\citep{2002AJ....124.1810S} with $0.1 < z < 0.2$ and K+e corrected all
of them to $z$ = 0.2. The SDSS main sample consists of all galaxies
with $r <$ 17.77 and provides a representative sample of the whole
galaxy population. For this sample, we generate the K+e corrections by
interpolating between a passive model and one with continuous star
formation based on the observed $g-i$ colour of the galaxy. We then
applied the SDSS and 2SLAQ LRG selection criteria at the two redshifts
respectively.  We plot in Figure \ref{fig:selectioncmp} the
colour--magnitude relation ($^{0.2}(g-i)$ versus
M$_{^{0.2}r}$) for these galaxies, where $^{0.2}g$ is the $g$ filter
at $z$ = 0.2, illustrating which galaxies would be selected by each criteria
and their combination.  This figure clearly illustrates the bluer
2SLAQ selection as well as the magnitude dependence of the SDSS LRG colour selection (E01) as the selection cuts through the red sequence starting at
M$_{^{0.2}r} >$ -22.8.

Although the detailed colour evolution of LRGs remains unknown, we
will proceed by assuming a simple passive model for LRGs evolution and
thus make a self--consistent colour selection at both $z=0.2$
and $z=0.55$. We can then check whether the observed LF
evolution is consistent with this simple hypothesis and whether any
further evolution (beyond passive) is required to explain our
observations. To do this, we first use the K+e corrections described
above to correct all the LRGs in both samples to the redshift of the
other sample $i.e.$ we correct the 2SLAQ LRGs to $z$ = 0.2 and the
SDSS LRGs to $z$ = 0.55. We then require that for any individual LRG
to be included in our analysis of the LRG LF it must satisfy both the
SDSS criteria (at $z$ = 0.2) and the 2SLAQ criteria (at
$z$ = 0.55). As might be expected, considering the broader
colour and magnitude ranges of the 2SLAQ selection criteria, most of
the SDSS LRGs ($\simeq$90\%) would still be selected as LRGs at $z$ =
0.55. The opposite is not true with only $\simeq$30\% of the 2SLAQ
LRGs satisfying the stricter SDSS Cut I LRG criteria.

\begin{figure*}
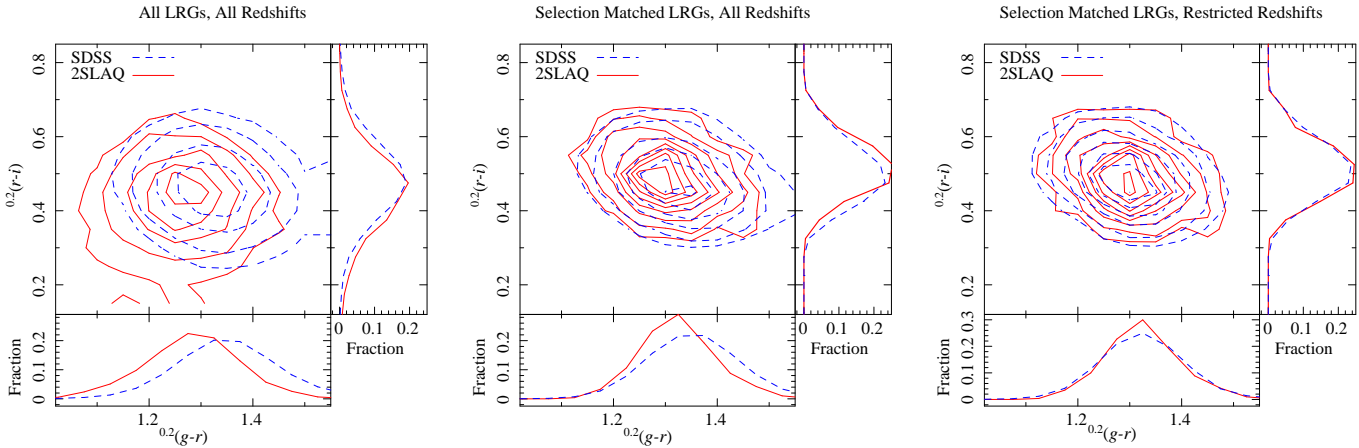

\vspace{6.5cm}
\includegraphics{colourdistoverlapcontourpaper2_nozcut_1.eps}
\includegraphics{colourdistoverlapcontourpaper2_nozcut_2.eps}
\includegraphics{colourdistoverlapcontourpaper2_2.eps}
\caption{\label{fig:coldistcont} The $g-r$ and $r-i$ colours of all the SDSS and 2SLAQ LRGs K+e corrected to $z$ = 0.2. The SDSS distributions have been convolved with the typical 2SLAQ photometric errors. The left--hand plot shows all the LRGs in these two samples, while the middle plot shows only those LRGs (in both samples) that match both the SDSS and 2SLAQ LRG selection criteria (described in the text). The right--hand plot shows those LRGs that match both the SDSS and 2SLAQ LRG selection criteria and have an additional redshift restriction; $0.17<z<0.24$ for the SDSS LRGs and $0.5<z<0.6$ for the 2SLAQ LRGs.}
\end{figure*}

Figure \ref{fig:coldistcont} shows the colour distributions of all the
2SLAQ LRGs ({\it red}) and all the SDSS LRGs ({\it blue}) convolved
with the typical photometric errors of the 2SLAQ LRGs, K+e corrected
to $z$ = 0.2.  The left--hand panel shows the colour distributions of
the full sample of SDSS and 2SLAQ LRGs considered in this paper (i.e.,
the raw data), while the middle panel shows only those galaxies which
satisfy both the SDSS and 2SLAQ LRG selection criteria at $z$ = 0.2
and $z$ = 0.55 respectively.  As mentioned previously, the 2SLAQ LRGs
in the left--hand panel have bluer colour distributions compared to
the SDSS LRGs, reflecting their bluer and fainter selection criteria.
When the joint SDSS and 2SLAQ selection criteria are applied, the
colour distributions of this restricted set of 2SLAQ LRGs now become
redder and much closer to the SDSS LRG colour distributions. However,
an offset between the 2SLAQ and SDSS LRG colour distributions is still
evident (in the middle panel), i.e., 0.05 magnitudes in the mean $g-r$
colour of LRGs. We believe this offset is due to residual inaccuracies
in the models used for our K-corrections, although we have tried to
reduce this problem by correcting between overlapping filters. The
wide redshift range of our samples (particularly in the SDSS) results
in the errors in our K-corrections still being significant and
detectable.

To improve the agreement between the 2SLAQ and SDSS LRG samples, we
restrict the redshift ranges of these two samples to be closer to the
redshifts where the SDSS filters have their greatest overlap (Figure
2). We restrict the SDSS LRG sample to $0.17 < z < 0.24$ and 2SLAQ LRG
sample to $0.5 < z < 0.6$, which reduces the amplitude of the
K-corrections at z = 0.2 for the SDSS LRGs to a range of 0.005 to
0.06, compared to $0.01$ to $0.11$ without this restricted redshift
range. Likewise, the amplitude of the 2SLAQ LRG K-corrections reduce
to a range of 0.08 to 0.30, compared to $0.11$ to $0.36$ without the
restricted redshift range.  The right--hand panel of Figure
\ref{fig:coldistcont} shows the colour distributions of LRGs that
satisfy both the SDSS and 2SLAQ selection criteria within the
restricted redshift ranges discussed above. This results in the colour
distributions of the two LRG samples being almost identical; the
median colour difference is now less than 0.01 magnitudes. These
restricted redshift LRG samples give us greater confidence that, under
the assumption of passive evolution, we are now selecting the same
type of galaxy in both the SDSS and 2SLAQ surveys.  It also suggests
that the discrepancy between the observed and model colours is only
affecting the K-corrections and not the evolution corrections. The
final redshift--restricted samples contain 6326 SDSS LRGs, with $0.17
< z < 0.24$, and 1725 2SLAQ LRGs, with $0.5 < z < 0.6$.  Clearly
making such tight redshift cuts has resulted in a significant
reduction in the size of our samples.  However, we believe that
minimising the errors in the K+e corrections is vital to ensure that
the LRG samples from the SDSS and 2SLAQ are as close as possible.

\begin{table}
  \begin{center}
    \caption{\label{tab:photerr}The median photometric errors for the SDSS and 2SLAQ LRGs from the single--epoch SDSS photometry and the multi-epoch SDSS photometry described in the text.}
    \begin{tabular}{c  c  c  c  c  c} 
      \multicolumn{1}{c}{} &
      \multicolumn{1}{c}{$u$} &
      \multicolumn{1}{c}{$g$} &
      \multicolumn{1}{c}{$r$} &
      \multicolumn{1}{c}{$i$} &
      \multicolumn{1}{c}{$z$} \\
      \hline \hline 
        SDSS single-epoch & 0.56 & 0.04 & 0.01 & 0.01 &  \\
        2SLAQ single-epoch &  & 0.15 & 0.05 & 0.03 & 0.09 \\
        2SLAQ multi-epoch  &  & 0.05 & 0.02 & 0.01 & 0.03 \\
      \hline
    \end{tabular}
  \end{center}
\end{table}

\subsection{Photometric errors}
\label{sec:phot}

Another potential bias that could affect the sample selection is the
larger photometric errors on the 2SLAQ LRGs compared to the SDSS LRGs
(because they are intrinsically fainter and from the same imaging dataset).
  This could systematically
change the selection, as a function of redshift, as more 2SLAQ LRGs
could be scattered both in and out of the sample as the photometric
error increases.

\begin{figure*}
\vspace{6.0cm}
\includegraphics{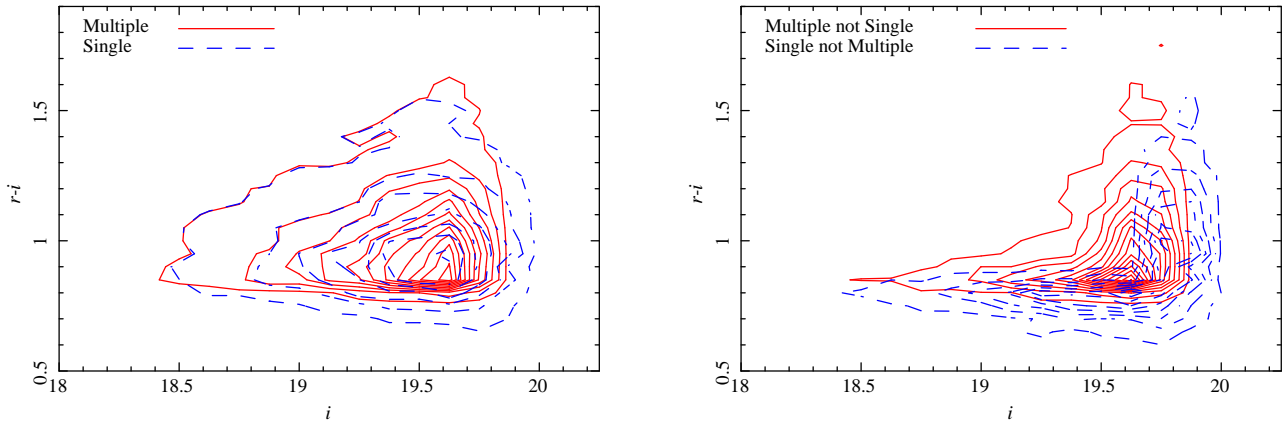}
\caption{\label{fig:coldistcoaddcont} The colour--magnitude relationship for LRGs satisfying the 2SLAQ selection criteria. We only show here LRGs within the 2SLAQ survey area covered by the SDSS multi-epoch photometry. The magnitudes used here were taken from the multi--epoch photometry as they are better signal--to--noise (Table 1). The left--hand panel shows LRGs that satisfy the 2SLAQ selection criteria for the single--epoch (dashed contours) and multi--epoch photometry (solid contours). The right--hand panel shows LRGs only selected in one of the two sets of photometry; red solid contours are LRGs selected in the multi--epoch photometry but missed in the single--epoch photometry, and visa-versa for the blue dashed contours.}
\end{figure*}

We attempt to measure this effect by utilising the multi-epoch SDSS
imaging data available over a subsample of the 2SLAQ survey area
\citep[see ][]{2005MNRAS.358..441B,2005scranton}.  This multi-epoch
data covers a total of $\simeq$190 deg$^2$ of the southern part of the
2SLAQ survey, and provides better signal--to--noise photometry for
LRGs in this area as demonstrated by Table \ref{tab:photerr}. We begin
by comparing the number of LRGs that satisfy the 2SLAQ selection
criteria using both the single and multi-epoch photometry. We find
that 10059 2SLAQ LRG targets were selected from the single-epoch
photometry compared to 10265 2SLAQ LRG targets selected using the
multi-epoch data. However, 25\% of these targets are different between
the two samples and have been scattered across the selection
boundaries in almost equal numbers.  In Figure
\ref{fig:coldistcoaddcont}, we show the colour--magnitude relationship
for LRGs both in common between the multi and single-epoch photometry
as well as the LRGs only selected in one of the two data sets. As
expected, LRGs selected only in the single--epoch data (i.e., missed
by the 2SLAQ selection using the multi--epoch photometry) are fainter
and bluer than 2SLAQ LRGs scattered out of the single--epoch selection
but selected using the multi--epoch photometry (the right-hand panel
of Figure 5). We note however that the colour--magnitude relationship
of LRGs for both the multi and single-epoch photometry is nearly
identical for $i <$ 19.3. This magnitude corresponds to
M$_{^{0.2}r}$ = -23.0 at $z$ = 0.6, the upper redshift limit used
herein to calculate the luminosity function.

\begin{figure}
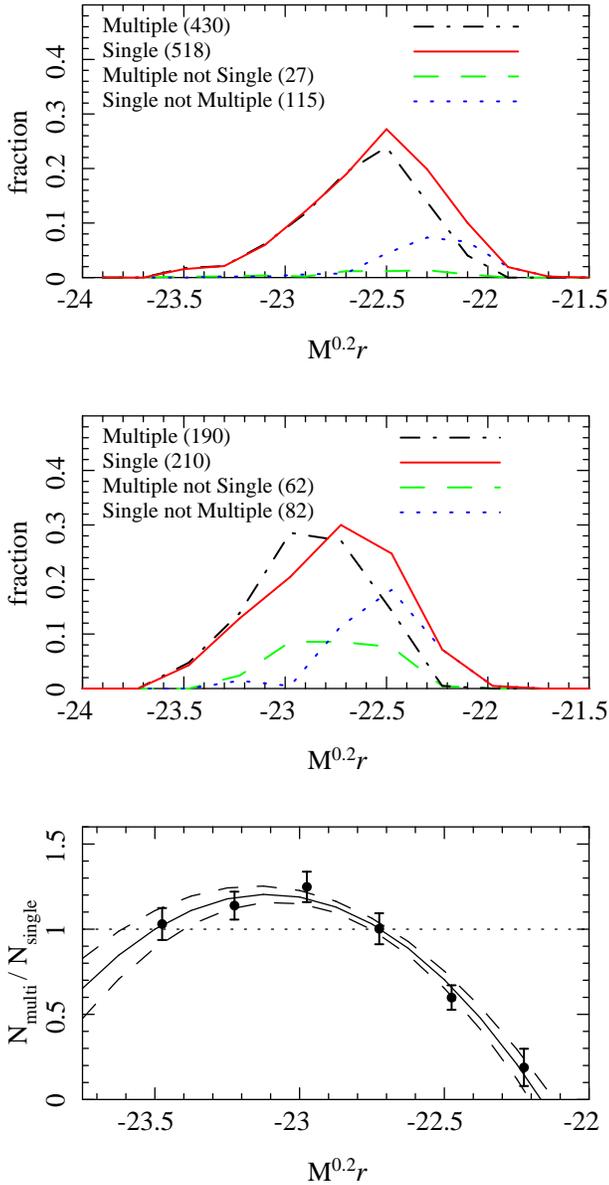

  \vspace{16.0cm} 
  \includegraphics{selectionspecdistspaper_mag.ps}
  \includegraphics{selectionspecdistspaper3_mag.ps}
 \includegraphics{selectionspecdistsratio_monti_mag.ps}  
\caption{\label{fig:coldistcoaddspec} Top and Middle - The magnitude distributions for all 2SLAQ LRGs in the redshift range $0.5 < z < 0.6$ with multi-epoch photometry. This subsample is further split into LRGs selected from the multi-epoch data (dot-dashed line), from the single-epoch data (solid line), and those LRGs selected only from multi--epoch data but not the single epoch data (dashed line) and visa-versa (dotted line). The distributions are normalised to the number in the single-epoch subsample. The number of LRGS in each of these subsamples is given in parentheses in the top panel. The top panel shows those passing the original 2SLAQ selection criteria, the middle those passing the additional selection criteria discussed in Section \ref{sec:sample}. Bottom - The ratio of number selected using the multi-epoch data to the number selected using the single epoch data using the additional selection criteria discussed in Section \ref{sec:sample}. The solid line shows a polynomial fit and the dashed lines the 1 sigma errors on the fit.}
\end{figure}

As a further test, we can limit this comparison, to only LRGs with measured 
redshifts. This is possible
because a subset of the 2dF fibres ($\sim$ 30\%) were allocated to
galaxy targets that lie slightly beyond the original 2SLAQ colour
selection boundaries for the highest priority ``Sample 8'' selection
\citep[for details see][]{2006Cannon}.  These extra LRG redshifts
allow us to investigate the effect of photometric errors on the
completeness of ``Sample 8'', although we are limited by the small 
size of the colour region beyond sample 8 when considering those 
galaxies that weren't selected but should have been. In Figure 
\ref{fig:coldistcoaddspec} (top), we
show the multi-epoch colour and magnitude distributions for 2SLAQ
LRGs, in the redshift range $0.5<z<0.6$, all K+e corrected to $z$ =
0.2 as before. As above, 22\% of the 2SLAQ LRGs are scattered into the
sample because of photometric errors, i.e., they satisfy the selection
criteria in the single--epoch photometry, but fail the criteria for
the multi--epoch photometry. Again, these LRGs are bluer and fainter
(in absolute magnitude) as shown by the dotted blue line in Figure
\ref{fig:coldistcoaddspec} (top). However, brighter than M$_{^{0.2}r} <$ = -22.65
the magnitude distributions of those selected by the single and multi-epoch 
data are almost identical.  
We are therefore confident that the
photometric errors have minimal effect on the 2SLAQ sample
function brighter than this limit. 

A more significant effect of the photometric errors occurs when we apply the 
additional selection criteria we use to match the samples as discussed in Section 
\ref{sec:sample}. We can again try to quantify this effect using the multi-epoch data
 and again show in Figure \ref{fig:coldistcoaddspec} (centre) the magnitude distributions
 of the single and multi-epoch selected 2SLAQ LRGs. However, here we only show those 
LRGs in the redshift range 0.5 $<$ z $<$ 0.6 that pass both both the SDSS and 2SLAQ selection 
criteria. 
Unlike previously, where we had a
 limited set of spectra for galaxies beyond the selection boundaries, in this instance 
we have many galaxies far beyond the selection boundaries due to the much bluer and fainter 
2SLAQ selection. This is clearly visible when comparing the second and fourth panels of 
Figure \ref{fig:selectioncmp}. On inspection of the magnitude distributions in Figure 
\ref{fig:coldistcoaddspec} it is clear that 
a significant fraction of galaxies are being scattered across the selection boundaries
 even for LRGs as bright as M$_{^{0.2}r}$ = -23.3. This is more clearly illustrated by 
the bottom panel of Figure \ref{fig:coldistcoaddspec} where we plot the ratio of the number of
 LRGs selected using the multi-epoch data to the number selected using the single-epoch data as 
a function of absolute magnitude. Since we are confident that we have fully sampled the 
colour-magnitude space beyond the selection boundaries we can use this ratio to correct 
the luminosity function and integrated number and luminosity densities presented in the 
Section \ref{sec:lf}. We note that the only place that we are not sampling beyond the 
boundary is for the faintest and reddest objects and we are thus not confident of the 
correction and any resulting quantities for M$_{^{0.2}r} >$ -22.4. 
In order to generate accurate errors on this correction, we take the 
multi-epoch selected sample and add random Gaussian errors typical of the 2SLAQ 
photometric errors to each magnitude. We then calculate how many LRGs would be selected 
in our final sample. We repeat this procedure 10000 times and measure the standard deviations 
in each bin which are shown as the errors in Figure \ref{fig:coldistcoaddspec} (bottom).

So far we have only discussed the effect of the photometric errors on the 
2SLAQ LRGs. It is also worth briefly discussing any potential effect on the SDSS LRGs.
As listed in Table \ref{tab:photerr} the typical photometric errors on the SDSS LRGs are 
much smaller than those of 2SLAQ and comparable to the multi-epoch errors on the 2SLAQ LRGs, 
except in the case of the $u$ band. However, the $u$ band data are only used when applying 
the 2SLAQ selection criteria to the SDSS LRGs and since the 2SLAQ criteria are typically 
significantly bluer and fainter than the SDSS criteria one would expect a small effect. In 
fact only 5\% of the SDSS LRGs are removed when the 2SLAQ criteria are applied to them 
and most of 
these (3\%) are as a result of the 2SLAQ magnitude limits where the $r$ band magnitude is 
used. We are therefore confident that the photometric errors on the SDSS LRGs result in an 
insignificant amount of scattering across the selection boundaries. 

\section{Luminosity Functions}

\subsection{Redshift Completeness}
\label{sec:comp}

The 2SLAQ survey is not spectroscopically complete unlike the SDSS LRG
sample which is $>99\%$ complete for input target redshifts.
Therefore, we must correct the 2SLAQ survey for this redshift
incompleteness, taking into account any dependence on magnitude and/or
colour. The redshift completeness is defined as the ratio of the number 
of Sample 8 2SLAQ LRGs with a reliable redshift to the number of Sample 
8 target LRGs selected from the SDSS imaging in each observed 2dF field.
To calculate this, we need to define the exact
survey area covered by the 2SLAQ survey in order to determine the
number of possible LRG targets. To achieve this, we repeatedly ran the
2dF configuration software on random positions until we had configured
over 5 million random points in a single 2dF field.  This exercise
provides a detailed map of all possible positions available to the 2dF
fibres within the field of view. We then built a random catalogue for
the whole 2SLAQ survey area by placing this single randomised 2dF
field at every observed field centre (see Cannon et al. 2006).
Finally, we remove any regions not in the original target input
catalogue, i.e., edges of the 2dF fields, that extended beyond the SDSS
photometry, and holes in the SDSS coverage. This produced a random
catalogue of approximately 400 million positions, covering every
possible position a 2dF fibre could have been placed throughout the
whole 2SLAQ survey. We then pixelised these random positions (into 30
by 30 arcsecond pixels) to generate a survey mask and positively
flagged all pixels that contain at least one random position.

\begin{figure}
\vspace{16.0cm}
\includegraphics{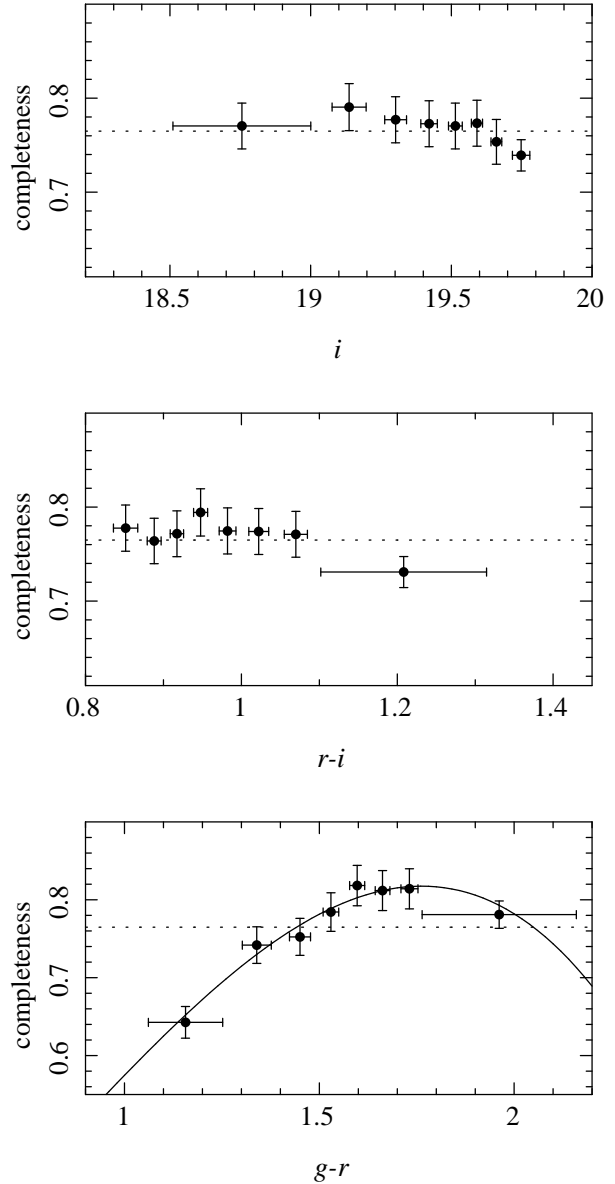}
\caption{\label{fig:comp} The redshift completeness of 2SLAQ LRGs as a function of apparent $i$ magnitude (top panel), $r-i$ colour (middle panel) and $g-r$ colour (bottom panel). The solid line in the bottom panel shows the 3rd order polynomial fit to the data and used to correct the sample for this incompleteness as a function of $g-r$ colour. The bins, chosen to contain 800 LRGs each, are plotted at the mean magnitude of the bin with the one sigma error bars.}
\end{figure}

The survey mask was used in two ways. First, we calculated the area of
the survey by summing all positive pixels, giving an area of 180.03
deg$^2$. Secondly, we used the mask to define those LRGs in the input
catalogue that could have been included in the 2SLAQ survey in order
to calculate the redshift completeness. We also restrict the observed
LRGs in the same manner resulting in about $0.5$\% 
of the observed LRGs (with redshifts) being excluded. This is caused by 
slight changes made to the 2dF configuration software during the 2SLAQ 
survey which we are unable to account for when constructing our mask. 
Figure \ref{fig:comp} shows the redshift completeness of
the 2SLAQ survey as a function of magnitude and colour. There are no
significant dependence of the redshift completeness on the $r-i$ colour, 
and $i$ magnitude only shows any significant dependence fainter than $i$ = 19.7.
However, we do witness a dependency on the
$g-r$ colour of the LRGs.  We correct for this dependence by fitting a
3rd order polynomial to the data (as shown in Figure \ref{fig:comp}) and use this
function to calculate the completeness for each LRG depending on its
observed $g-r$ colour. Including this correction changes the 2SLAQ LRG
LF by $<$ 1\% compared to assuming an overall redshift completeness of
76.5\% regardless of its $g-r$ colour.

\subsection{The Luminosity Function}
\label{sec:lf}
\begin{figure}
  \vspace{11.5cm} \includegraphics{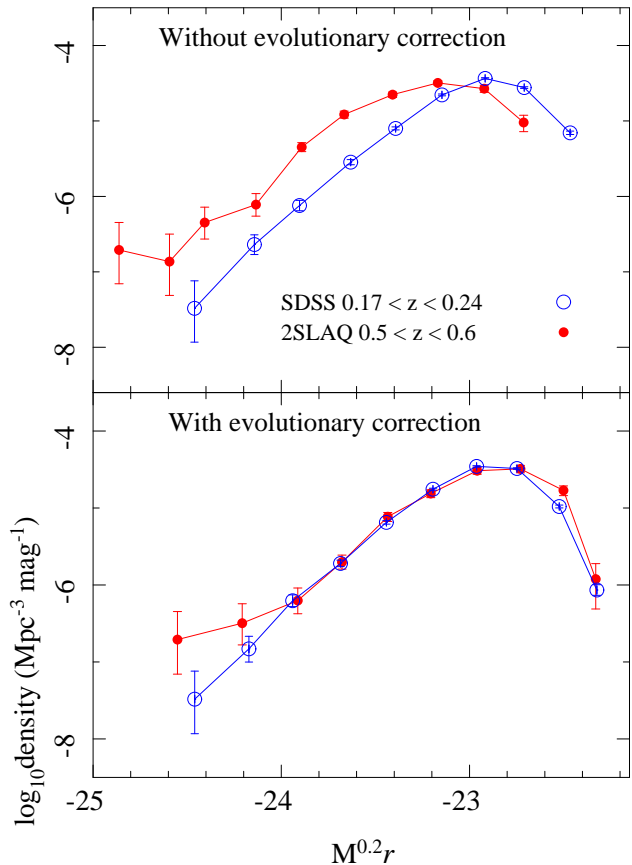}
\caption{\label{fig:lumfn} The M$_{^{0.2}r}$ luminosity function without (top panel) and with (bottom panel) passive evolution corrections for both the SDSS (open data points) and 2SLAQ (solid data points) LRG samples. The points are plotted with their one sigma error bars as described in the text.}
\end{figure}

\begin{table}
  \begin{center}
    \caption{\label{tab:LFE}The M$_{^{0.2}r}$ luminosity function with evolutionary corrections for the SDSS and 2SLAQ LRGs. The values of M$_{^{0.2}r}$ are the median of the bins, the density is given in units of 10$^{-6}$ Mpc$^{-3}$ mag$^{-1}$ and the errors are 1 sigma.}
    \begin{tabular}{c  c  c  c} 
      \multicolumn{2}{c}{SDSS} &
      \multicolumn{2}{c}{2SLAQ} \\
      \multicolumn{1}{c}{M$_{^{0.2}r}$}&
      \multicolumn{1}{c}{Density} &
      \multicolumn{1}{c}{M$_{^{0.2}r}$} &
      \multicolumn{1}{c}{Density} \\
      \hline \hline 
        -24.45  &  0.03 $\pm$  0.03 & -24.55 &  0.20 $\pm$  0.19\\
        -24.17  &  0.15 $\pm$  0.06 & -24.20 &  0.32 $\pm$  0.20\\
        -23.94  &  0.62 $\pm$  0.11 & -23.91 &  0.63 $\pm$  0.24\\
        -23.68  &  1.91 $\pm$  0.19 & -23.68 &  1.99 $\pm$  0.42\\
        -23.44  &  6.51 $\pm$  0.34 & -23.43 &  7.72 $\pm$  1.03\\
        -23.19  & 17.56 $\pm$  0.55 & -23.21 & 15.45 $\pm$  1.81\\
        -22.96  & 34.47 $\pm$  0.76 & -22.96 & 30.59 $\pm$  3.11\\
	-22.75  & 32.59 $\pm$  0.75 & -22.73 & 32.25 $\pm$  3.59\\
	-22.52  & 10.47 $\pm$  0.44 & -22.50 & 16.94 $\pm$  2.44\\
	-22.32  & 0.86  $\pm$  0.14 & -22.33 & 1.19  $\pm$  0.70\\
      \hline
    \end{tabular}
  \end{center}
\end{table}

\begin{table}
  \begin{center}
    \caption{\label{tab:LFNE}The M$_{^{0.2}r}$ luminosity function without evolutionary corrections for the SDSS and 2SLAQ LRGs. The values of M$_{^{0.2}r}$ are the median of the bins, the density is given in units of 10$^{-6}$ Mpc$^{-3}$ mag$^{-1}$ and the errors are 1 sigma.}
    \begin{tabular}{c  c  c  c} 
      \multicolumn{2}{c}{SDSS} &
      \multicolumn{2}{c}{2SLAQ} \\
      \multicolumn{1}{c}{M$_{^{0.2}r}$}&
      \multicolumn{1}{c}{Density} &
      \multicolumn{1}{c}{M$_{^{0.2}r}$} &
      \multicolumn{1}{c}{Density} \\
      \hline \hline 
         --  &  --                  & -24.86 &  0.20 $\pm$  0.19\\
         --  &  --                  & -24.59 &  0.14 $\pm$  0.13\\
        -24.46  &  0.03 $\pm$  0.03 & -24.41 &  0.45 $\pm$  0.22\\
        -24.14  &  0.23 $\pm$  0.07 & -24.13 &  0.78 $\pm$  0.27\\
        -23.90  &  0.76 $\pm$  0.12 & -23.89 &  4.50 $\pm$  0.62\\
        -23.63  &  2.84 $\pm$  0.22 & -23.67 & 12.18 $\pm$  1.30\\
        -23.39  &  7.96 $\pm$  0.37 & -23.41 & 22.29 $\pm$  1.83\\
        -23.15  & 22.15 $\pm$  0.61 & -23.17 & 31.95 $\pm$  2.05\\
        -22.92  & 36.64 $\pm$  0.78 & -22.92 & 26.79 $\pm$  2.88 \\
	-22.71  & 27.68 $\pm$  0.69 & -22.71 &  9.54 $\pm$  2.34 \\
	-22.47  & 6.96  $\pm$  0.36 &   --   &        --         \\

     \hline
    \end{tabular}
  \end{center}
\end{table}

In Figure \ref{fig:lumfn}, we present the M$_{^{0.2}r}$ luminosity
function (LF) of the 2SLAQ and SDSS LRG samples as described above,
both without (top panel) and with (bottom panel) passive evolution
corrections.  We calculated these LFs using the 1/V$_{\rm max}$
method, where for each galaxy we compute the maximum and minimum
redshifts at which it would have been selected, including the K+e
corrections described above. For all LRGs brighter than M$_{^{0.2}r}$
= -23, the maximum and minimum redshifts correspond to redshift limits
described previously, namely $0.17 < z < 0.24$ for the SDSS sample and $0.5
< z < 0.6$ for the 2SLAQ sample. Therefore, above this absolute
magnitude, both samples are essentially volume--limited. The
luminosity functions were then determined by calculating the volume
(V$_{\rm max}$) within which each galaxy is detectable, modified by the
colour--dependent redshift completeness corrections for the 2SLAQ
sample, and summed over all galaxies in the sample. We then apply the photometric error 
scattering corrections discussed in Section \ref{sec:phot} to each bin. We provide the
numerical values of the SDSS and 2SLAQ LFs in Tables \ref{tab:LFE} and
\ref{tab:LFNE}.

We use jack--knife (JK) re--sampling to calculate the errors on our
luminosity functions. This is achieved by splitting the SDSS and 2SLAQ
samples into 20 subregions, of equal area, and re--calculating 20
LFs with each of these subregions removed in turn. We find that for the
2SLAQ sample the JK errors are up to 30\% larger than the Poisson
errors in the four faintest magnitude bins (those which contain the
most galaxies), while the two error estimates are the same for the
brighter bins. We therefore quote in Tables \ref{tab:LFE} and
\ref{tab:LFNE}, and plot in Figure \ref{fig:lumfn}, the larger of
these two errors combined with the errors introduced by the photometric 
error scattering correction. For the SDSS sample, we find that the JK errors are
consistent with the Poisson errors for all magnitude bins; we
therefore use the Poisson errors.

\begin{table*}
  \begin{center}
    \caption{The integrated number and luminosity density of the evolution corrected SDSS and 2SLAQ samples and the ratio of these two measurements.\label{tab:den}}
    \begin{tabular}{c  c  c  c  c  c  c  } 
      \multicolumn{1}{c}{Sample} &
      \multicolumn{3}{c}{Density ($\times 10^{-6}$ Mpc$^{-3}$)} &
      \multicolumn{3}{c}{Luminosity Density ($\times 10^{6}$ L$_{\odot}$ Mpc$^{-3}$)}\\
      \multicolumn{1}{c}{} &
      \multicolumn{1}{c}{M$_{^{0.2}r}$ $<$ -22.5} &
      \multicolumn{1}{c}{M$_{^{0.2}r}$ $<$ -23.0} &
      \multicolumn{1}{c}{M$_{^{0.2}r}$ $<$ -23.5} &
      \multicolumn{1}{c}{M$_{^{0.2}r}$ $<$ -22.5} &
      \multicolumn{1}{c}{M$_{^{0.2}r}$ $<$ -23.0} &
      \multicolumn{1}{c}{M$_{^{0.2}r}$ $<$ -23.5} \\
      \hline \hline 
        SDSS  & 25.09 $\pm$ 0.33  & 9.77 $\pm$ 0.20 & 1.12 $\pm$ 0.07 & 3.62 $\pm$ 0.05 & 1.79 $\pm$ 0.04 & 0.31 $\pm$ 0.02 \\
        2SLAQ & 24.36 $\pm$ 1.47 & 9.33 $\pm$ 0.68 & 1.24 $\pm$ 0.16 & 3.53 $\pm$ 0.21 & 1.78 $\pm$ 0.12 & 0.36 $\pm$ 0.05 \\
      \hline
        Ratio & 1.03 $\pm$ 0.06  & 1.05 $\pm$ 0.08 & 0.91 $\pm$ 0.13 & 1.03 $\pm$ 0.06 & 1.01 $\pm$ 0.08 & 0.86 $\pm$ 0.13  \\
        \hline  
    \end{tabular}
  \end{center}
\end{table*}

Table \ref{tab:den} lists the integrated number and luminosity densities for a series of 
magnitude limits for the 2SLAQ and SDSS samples and their ratio. These are again corrected for the effect of the photometric errors scattering galaxies into and out of the sample. However, in this instance we use the fit to the relation shown in Figure \ref{fig:coldistcoaddspec} as we need to make a correction to each individual galaxy. The errors are again a combination of the JK errors and those from the photometric scattering.

\section{Discussion}
\label{sec:dis}

As seen in Figure \ref{fig:lumfn}, the SDSS and 2SLAQ LFs brighter
than M$_{^{0.2}r}$ = -22.6 are in excellent agreement when
the passive evolution corrections are included. Fainter than this limit we
 are not confident of the photometric scattering correction we have made, although 
we note that the LFs are still in reasonable agreement. 
The agreement of these luminosity functions is further confirmed by
calculating the integrated number and luminosity density of LRGs as
given in Table \ref{tab:den}.  Brighter than M$_{^{0.2}r}$ = -22.5, both
the integrated luminosity and number density of the 2SLAQ and SDSS
samples agree to within their one sigma errors, and are measured to
better than 10\% out to $z=0.6$.

Throughout the analysis presented herein, we have consistently used
the same simple passive evolution model for predicting and correcting
the colours and luminosities of LRGs as a function of redshift, and
this agreement demonstrates the lack of any extra evolution, beyond
the passive fading of old stars, out to $z \simeq 0.6$. This result
confirms the underlying assumptions of \citet{2001AJ....122.2267E},
and \citet{2006Cannon}, that the majority of LRGs out to $z\simeq0.6$
can be selected via straightforward colour cuts, in multicolour data,
assuming simple passive evolution of their stellar populations. The
result also confirms the work of
\citet{2003AJ....125.1866B,2003AJ....125.1882B} for lower redshift
massive ellipticals in the SDSS.

It may appear that our lack of extra evolution beyond passive (out to
$z\sim0.6$) is in conflict with recent results from the C17, DEEP2, and
SXDS surveys \citep{2004ApJ...608..752B,Faber,2005ApJ...634..861Y}.
These smaller--area, but deeper (in magnitude limit and redshift),
surveys find evidence for a change in the density of red galaxies out
to $z\sim1$ beyond that expected from passive fading of the stellar
populations. For example, \citet{Faber} report a quadrupling of
$\phi^*$ for red galaxies since $z=1$, although this result is
strongest in their highest redshift bin, where they admit their data
are weakest. A direct comparison with these deeper surveys is
difficult because of the differences in colour selections used for the
surveys, as well as the relative luminosity ranges probed by the
different surveys, i.e., the 2SLAQ survey is designed to probe
galaxies brighter than a few $L^*$, while the DEEP2, SXDS and C17
surveys effectively probe galaxies below $L^*$ at $z\sim0.6$ (due to
their smaller areal coverage and fainter magnitude limits).

\begin{figure}
\vspace{6.5cm}
\includegraphics{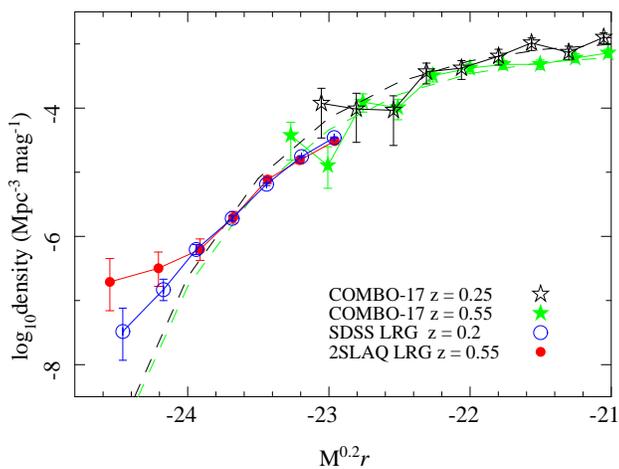}
\caption{\label{fig:lumfncombo} The M$_{^{0.2}r}$ luminosity function with passive evolution corrections for the SDSS (blue open data points), 2SLAQ (red solid data points) LRG samples, and the COMBO-17 red galaxies at z = 0.25 (black open stars) and z = 0.55 (green solid stars) \citep{2004ApJ...608..752B}. The dashed lines show the Schechter function fit to the COMBO-17 points. The points are plotted with their one sigma errors.}
\end{figure}

However, to facilitate such a comparison, we show in Figure
\ref{fig:lumfncombo}, the LFs from Figure
\ref{fig:lumfn}, and the C17 red galaxy LFs \citep[Figure
3]{2004ApJ...608..752B} for the same redshift range and K+e corrected
to M$_{^{0.2}r}$. We only plot our LFs to M$_{^{0.2}r} <$ -22.9 as we do 
not include all the red galaxies fainter than this due to the SDSS LRG 
selection criteria.
 Figure \ref{fig:lumfncombo} demonstrates that when 
one restricts the
data to the same redshift range, there is excellent qualitative
agreement between the 2SLAQ and C17 luminosity functions. We are unable to 
make a quantitative comparison due to the difficulty in exactly matching the 
selection criteria of the two surveys.
 Taken
together, the surveys shown in Figure \ref{fig:lumfncombo} extend the 
evidence for no evolution in the LF of
LRGs to M$_{^{0.2}r} < -21$, which is close to $L^{\star}$ in the LF.
Figure \ref{fig:lumfncombo} also demonstrates that these two surveys 
are probing
different luminosity regimes at $z<0.6$ as there is at most only 0.5
magnitudes of overlap in their LFs in which the C17 survey is becoming seriously affected 
by small
number statistics due to its smaller areal coverage.

\begin{figure}
  \vspace{8.5cm} \includegraphics{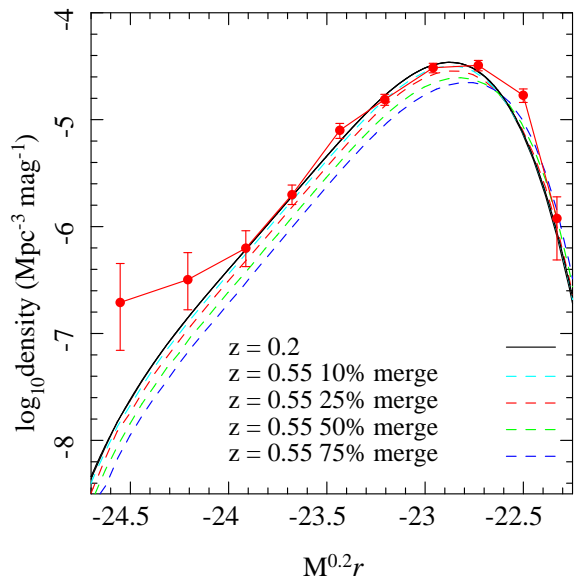}
\caption{\label{fig:lumfnmerging} The M$_{^{0.2}r}$ luminosity function with passive evolution corrections for the 2SLAQ LRGs (solid data points) and fit to the SDSS LRGs black line. The lines show the effect of splitting varying fractions of the 2SLAQ LRGs in two, simulating major mergers between z = 0.6 and z = 0.2. }
\end{figure}

The C17 data presented in Figure 9 agrees with our findings that for
the brightest galaxies there appears to be no evidence for density
evolution out to $z\sim0.6$. This result is not necessarily in
conflict with the work of
\citet{2004ApJ...608..752B,Faber,2005ApJ...634..861Y}, as we are still
probing different redshift and luminosity ranges than these other
studies. Taken together, these results could indicate the existence of
different evolutionary scenarios above and below $L^*$ in the
luminosity function, i.e., above $L^*$, galaxies have only evolved
passively (since $z=0.6$), while below this luminosity, the red galaxy
population is experiencing significant evolution.
\citet{2004MNRAS.350.1005K} sees similar evidence for differential
evolutionary trends with luminosity at $z\sim1$, and claim this
supports the idea of ``down-sizing'' i.e., the galaxy evolutionary
processes (like star--formation and assembly) decrease with increased
luminosity as a function of redshift. These higher redshift
observations are also consistent with the observed transition, at
$M_r\sim-20.5$, in the local colour--magnitude relationship between a
dominant ``red'' population of galaxies (above this luminosity),
compared to ``blue'' population below. Likewise,
\citet{2003MNRAS.341...54K} find a significant change in the
distribution of stellar masses of local galaxies at the same
luminosity. A more detailed joint analysis of the 2SLAQ and deeper
surveys will be presented in a forthcoming paper.

\begin{table}
  \begin{center}
    \caption{The chi--squared values for fitting difference merger fractions\label{tab:chi}}
    \begin{tabular}{c  c  c  c  c} 
      \multicolumn{1}{c}{Fraction} &
      \multicolumn{2}{c}{M$_{^{0.2}r} <$ -22.75} &
      \multicolumn{2}{c}{M$_{^{0.2}r} <$ -23.25} \\
      \multicolumn{1}{c}{Merging} &
      \multicolumn{1}{c}{Reduced $\chi^2$} &
      \multicolumn{1}{c}{Prob}&
      \multicolumn{1}{c}{Reduced $\chi^2$} &
      \multicolumn{1}{c}{Prob}\\
      \hline \hline 
        0  & 0.68 & 0.70 & 0.53  & 0.76 \\
        0.10 & 0.64 & 0.74 & 0.90 & 0.47 \\
        0.25 & 1.66 & 0.10 & 2.00 & 0.07 \\
        0.50 & 4.54 & $<$ 0.0001 & 3.52 & 0.003  \\
        0.75 & 6.07 & $<$ 0.0001 & 6.08 & $<$ 0.0001  \\
      \hline
    \end{tabular}
  \end{center}
\end{table}

The luminosity functions given in Figure \ref{fig:lumfn} place tight
constraints on models of massive galaxy formation and evolution. Our
results appear to favour little, or no, density evolution, as we
only require the expected passive evolution of the luminosities of
these LRGs to explain the observed differences in their LFs as a
function of redshift. In other words, there are already enough LRGs
per unit volume at $z\simeq0.6$ to account for the density of LRGs
measured at $z\simeq 0.2$. To study this further, we must compare our
results with the latest predictions for massive galaxy evolution. For
example, \citet{2005astroph0509725} have used the effects of AGN
feedback to regulate new star--formation in massive ellipticals within
their semi--analytical Cold Dark Matter (CDM) model of galaxy
formation.  As shown in Figures 4 \& 5 of \citet{2005astroph0509725},
they find that 50\% of stars in $z=0$ massive ellipticals are already
formed by a median redshift of $z=2.6$, yet 50\% of the stellar mass
of $z=0$ massive ellipticals is not in place until a median redshift
of $z=0.8$.  In Figure 9 of their paper, they show that galaxies more
massive than $\simeq10^{11}{\rm M_{\odot}}$ are built up through $\sim
5$ major mergers, that must be ``dry'' (without gas) to prevent new
star-formation \citep{2005AJ....130.2647V}. Using the
\citet{2003MNRAS.344.1000B} models described earlier, we estimate that
our LRGs have stellar masses $> 5 \times 10^{11}{\rm M_{\odot}}$,
consistent with the massive galaxy sample discussed by
\citet{2005astroph0509725}.

We investigate a simple model, motivated by the idea of ``dry
mergers'' and the results of \citet{2005astroph0509725}, to simulate
the effect on the luminosity function of the hierarchical build--up of
these LRGs through major merger events. To achieve this, we fit the
SDSS LRG LF predict the higher redshift
(at $z=0.55$) 2SLAQ LRG LF under the assumption that a given fraction
of the SDSS LRGs were formed from a major merger of two equal mass
progenitors, i.e., we assume that two 2SLAQ LRGs have merged between
$z=0.55$ and $z=0.2$ to form a more massive SDSS LRG.  We then
determine the likely fraction of $z=0.2$ LRGs that could have been
formed this way by fitting (via $\chi^2$) this model to the observed
$z=0.55$ 2SLAQ LRG luminosity function.  We could have performed this
test using the actual data, rather than fitting the $z=0.2$ SDSS LRG
LF, but unfortunately such a method would suffer from small number
statistics at the bright end of the LF, resulting in the brightest bin
disappearing as there are no brighter LRGs being split to refill the
bin. However, the $\chi^2$ values are almost identical for the other
bins.

The results are presented in Figure \ref{fig:lumfnmerging} and listed
in Table \ref{tab:chi}. They reveal that the 2SLAQ and SDSS LFs are
consistent with each other without any need for merging. At the
3$\sigma$ level, we can exclude merger rates of $>50\%$, i.e., more
than half the LRGs at $z=0.2$ are already well-assembled, with more
than half their stellar mass in place, by $z\simeq0.6$. This
observation is consistent with \citet{masjedi} who find that LRG-LRG
mergers can not be responsible for the mass growth of LRGs at $z<0.36$
based on the small--scale clustering amplitude of SDSS LRG correlation
function.

Our limit is barely consistent with the predictions in Figure 5 of
\citet{2005astroph0509725}, where they show that $\sim 50\%$ of $z=0$
massive ellipticals have accreted 50\% of their stellar mass since
$z\simeq 0.8$. We note that our simple model does not constrain the
rate of minor mergers; the results of \citet{2006Roseboom} on the
spectral analysis of 2SLAQ LRGs suggests the $\simeq1\%$ of our LRGs
have experienced a small burst of star--formation in the last Gigayear
(based on the observed H$\delta$ line), which affects less than 10\%
of their stellar mass. 

Our results are a challenge for models of hierarchical galaxy
formation.  More detailed comparisons with semi--analytical CDM models
are required and will be investigated in other papers. For example,
\citet{2005astroph0511338} also include AGN feedback in their
semi-analytic simulations, but follow the model suggested by
\citet{2004MNRAS.347.1093B} whereby the AGN heating and the gas
cooling form a self--regulating feedback loop if the gas is in the
hydrostatic cooling regime (found in groups and cluster) and the
central black hole is suitably massive.  Initial results suggest that
the \citet{2005astroph0511338} prescription provides a better fit to
the LRG evolution discussed here (Bower, priv. comm.).  We also have
significantly more data than used in this paper, i.e., if we could
precisely model the K+e corrections of these LRGs over the joint
redshift range of the SDSS \& 2SLAQ surveys, we would gain a factor of
2 increase in the number of LRGs used to compute their luminosity
functions. In future work, we will investigate the use of other
stellar synthesis models for such corrections \citep{2005MNRAS.362..799M}.

\section*{Acknowledgements}

The authors are very grateful to Carlton Baugh, Richard Bower, Darren Croton, 
Yeong Loh, Claudia Maraston, Daniel Thomas, and Russell Smith for
advice and comments on this work.  The authors thank the AAO staff for
their assistance during the collection of these data. We are also
grateful to PPARC TAC and ATAC for their generous allocation of
telescope time to this project. DAW thanks the ICG Portsmouth for
their financial support during this work. RCN acknowledges the EU
Marie Curie program for their support. IRS and ACE acknowledge support from the 
Royal Society.

Funding for the SDSS and SDSS-II has been provided by the Alfred P.
Sloan Foundation, the Participating Institutions, the National Science
Foundation, the U.S. Department of Energy, the National Aeronautics
and Space Administration, the Japanese Monbukagakusho, the Max Planck
Society, and the Higher Education Funding Council for England. The
SDSS Web Site is http://www.sdss.org/.

The SDSS is managed by the Astrophysical Research Consortium for the
Participating Institutions. The Participating Institutions are the
American Museum of Natural History, Astrophysical Institute Potsdam,
University of Basel, Cambridge University, Case Western Reserve
University, University of Chicago, Drexel University, Fermilab, the
Institute for Advanced Study, the Japan Participation Group, Johns
Hopkins University, the Joint Institute for Nuclear Astrophysics, the
Kavli Institute for Particle Astrophysics and Cosmology, the Korean
Scientist Group, the Chinese Academy of Sciences (LAMOST), Los Alamos
National Laboratory, the Max-Planck-Institute for Astronomy (MPIA), the
Max-Planck-Institute for Astrophysics (MPA), New Mexico State
University, Ohio State University, University of Pittsburgh,
University of Portsmouth, Princeton University, the United States
Naval Observatory, and the University of Washington.

\appendix
\section{K and evolutionary corrections}
\label{sec:K+e}
We present here tables of the $g-i$ colours, K and K+e corrections derived
from the \citet{2003MNRAS.344.1000B} models described in Section
\ref{sec:KEcorr}.

\begin{table*}
  \begin{center}
    \caption{K and K+e corrections for the SDSS LRGs to z = 0.2 assuming the passive model discussed in the text}
    \begin{tabular}{ c  c  c  c  c  c  c  c  c  c} 
        \hline  
      \multicolumn{1}{c}{} &
      \multicolumn{1}{c}{} &
      \multicolumn{4}{c}{K} &
      \multicolumn{4}{c}{K+e} \\
      \multicolumn{1}{c}{z} &
      \multicolumn{1}{c}{$g-i$} &
      \multicolumn{1}{c}{$u$}&
      \multicolumn{1}{c}{$g$}&
      \multicolumn{1}{c}{$r$}&
      \multicolumn{1}{c}{$i$}&
      \multicolumn{1}{c}{$u$}&
      \multicolumn{1}{c}{$g$}&
      \multicolumn{1}{c}{$r$}&
      \multicolumn{1}{c}{$i$}\\
      \hline 
        0.150 & 1.697 & -0.398 & -0.280 & -0.086 & -0.035 & -0.312 & -0.214 & -0.030 &  0.017  \\
        0.175 & 1.813 & -0.207 & -0.141 & -0.043 & -0.018 & -0.162 & -0.108 & -0.015 &  0.008  \\
        0.200 & 1.929 &  0.000 &  0.000 &  0.000 &  0.000 &  0.000 &  0.000 &  0.000 &  0.000  \\
        0.225 & 2.034 &  0.202 &  0.132 &  0.042 &  0.017 &  0.159 &  0.098 &  0.015 & -0.008 \\ 
        0.250 & 2.121 &  0.422 &  0.254 &  0.088 &  0.043 &  0.336 &  0.185 &  0.032 & -0.008 \\ 
        0.275 & 2.208 &  0.647 &  0.376 &  0.140 &  0.069 &  0.523 &  0.272 &  0.056 & -0.007 \\ 
        0.300 & 2.300 &  0.879 &  0.506 &  0.190 &  0.097 &  0.726 &  0.367 &  0.079 & -0.005 \\
        0.325 & 2.393 &  1.116 &  0.644 &  0.242 &  0.128 &  0.934 &  0.466 &  0.104 &  0.001 \\ 
        0.350 & 2.450 &  1.333 &  0.753 &  0.303 &  0.168 &  1.125 &  0.537 &  0.137 &  0.015 \\ 
      \hline
    \end{tabular}
  \end{center}
\end{table*}

\begin{table*}
  \begin{center}
    \caption{K and K+e corrections for SDSS LRGs to z = 0.2 assuming the passive plus star--forming model discussed in the text}
    \begin{tabular}{ c  c  c  c  c  c  c  c  c  c} 
        \hline  
      \multicolumn{1}{c}{} &
      \multicolumn{1}{c}{} &
      \multicolumn{4}{c|}{K} &
      \multicolumn{4}{c}{K+e} \\
      \multicolumn{1}{c}{z} &
      \multicolumn{1}{c}{$g-i$} &
      \multicolumn{1}{c}{$u$}&
      \multicolumn{1}{c}{$g$}&
      \multicolumn{1}{c}{$r$}&
      \multicolumn{1}{c}{$i$}&
      \multicolumn{1}{c}{$u$}&
      \multicolumn{1}{c}{$g$}&
      \multicolumn{1}{c}{$r$}&
      \multicolumn{1}{c}{$i$}\\
      \hline 
        0.150 & 1.597 & -0.217 & -0.244 & -0.078 & -0.032 & -0.189 & -0.195 & -0.029 &  0.016  \\
        0.175 & 1.703 & -0.107 & -0.123 & -0.039 & -0.016 & -0.095 & -0.098 & -0.014 &  0.007  \\
        0.200 & 1.808 &  0.000 &  0.000 &  0.000 &  0.000 &  0.000 &  0.000 &  0.000 &  0.000  \\
        0.225 & 1.903 &  0.094 &  0.113 &  0.038 &  0.016 &  0.086 &  0.088 &  0.014 & -0.007 \\ 
        0.250 & 1.982 &  0.186 &  0.217 &  0.079 &  0.039 &  0.174 &  0.167 &  0.030 & -0.007 \\ 
        0.275 & 2.060 &  0.268 &  0.321 &  0.126 &  0.064 &  0.261 &  0.246 &  0.053 & -0.006 \\ 
        0.300 & 2.143 &  0.340 &  0.429 &  0.170 &  0.089 &  0.343 &  0.332 &  0.074 & -0.003 \\
        0.325 & 2.225 &  0.402 &  0.542 &  0.217 &  0.118 &  0.417 &  0.420 &  0.097 &  0.003 \\ 
        0.350 & 2.275 &  0.444 &  0.631 &  0.271 &  0.155 &  0.472 &  0.484 &  0.128 &  0.017 \\ 
      \hline
    \end{tabular}
  \end{center}
\end{table*}

\begin{table*}
  \begin{center}
    \caption{K and K+e corrections for 2SLAQ LRGs to z = 0.55 assuming the passive model discussed in the text}
    \begin{tabular}{ c  c  c  c  c  c  c  c  c  c} 
        \hline  
      \multicolumn{1}{c}{} &
      \multicolumn{1}{c}{} &
      \multicolumn{4}{c|}{K} &
      \multicolumn{4}{c}{K+e} \\
      \multicolumn{1}{c}{z} &
      \multicolumn{1}{c}{$g-i$} &
      \multicolumn{1}{c}{$g$}&
      \multicolumn{1}{c}{$r$}&
      \multicolumn{1}{c}{$i$}&
      \multicolumn{1}{c}{$z$}&
      \multicolumn{1}{c}{$g$}&
      \multicolumn{1}{c}{$r$}&
      \multicolumn{1}{c}{$i$}&
      \multicolumn{1}{c}{$z$}\\
      \hline 
        0.450 & 2.574 & -0.402 & -0.470 & -0.140 & -0.086 & -0.256 & -0.328 & -0.042 &  0.000 \\
        0.475 & 2.635 & -0.298 & -0.351 & -0.111 & -0.066 & -0.190 & -0.244 & -0.037 & -0.001 \\
        0.500 & 2.682 & -0.204 & -0.227 & -0.075 & -0.049 & -0.131 & -0.157 & -0.025 & -0.006 \\
        0.525 & 2.735 & -0.105 & -0.110 & -0.041 & -0.027 & -0.069 & -0.074 & -0.016 & -0.005 \\
        0.550 & 2.788 &  0.000 &  0.000 &  0.000 &  0.000 &  0.000 &  0.000 &  0.000 &  0.000 \\
        0.575 & 2.850 &  0.115 &  0.103 &  0.040 &  0.027 &  0.078 &  0.068 &  0.016 &  0.005 \\
        0.600 & 2.921 &  0.245 &  0.203 &  0.084 &  0.058 &  0.169 &  0.136 &  0.036 &  0.014 \\
        0.625 & 3.001 &  0.394 &  0.306 &  0.137 &  0.093 &  0.278 &  0.208 &  0.064 &  0.027 \\
        0.650 & 3.077 &  0.546 &  0.414 &  0.200 &  0.126 &  0.389 &  0.284 &  0.101 &  0.037 \\
      \hline
    \end{tabular}
  \end{center}
\end{table*}

\begin{table*}
  \begin{center}
    \caption{K and K+e corrections for 2SLAQ LRGs to z = 0.55 assuming the passive plus star-forming model discussed in the text.}
    \begin{tabular}{ c  c  c  c  c  c  c  c  c  c} 
        \hline  
      \multicolumn{1}{c}{} &
      \multicolumn{1}{c}{} &
      \multicolumn{4}{c|}{K} &
      \multicolumn{4}{c}{K+e} \\
      \multicolumn{1}{c}{z} &
      \multicolumn{1}{c}{$g-i$} &
      \multicolumn{1}{c}{$g$}&
      \multicolumn{1}{c}{$r$}&
      \multicolumn{1}{c}{$i$}&
      \multicolumn{1}{c}{$z$}&
      \multicolumn{1}{c}{$g$}&
      \multicolumn{1}{c}{$r$}&
      \multicolumn{1}{c}{$i$}&
      \multicolumn{1}{c}{$z$}\\
      \hline 
        0.450 & 2.376 & -0.244 & -0.412 & -0.126 & -0.080 & -0.189 & -0.308 & -0.039 & -0.001\\ 
        0.475 & 2.422 & -0.177 & -0.308 & -0.100 & -0.061 & -0.139 & -0.230 & -0.035 & -0.002\\ 
        0.500 & 2.455 & -0.118 & -0.199 & -0.067 & -0.045 & -0.095 & -0.148 & -0.024 & -0.006\\ 
        0.525 & 2.491 & -0.060 & -0.096 & -0.037 & -0.025 & -0.049 & -0.070 & -0.015 & -0.005\\ 
        0.550 & 2.526 &  0.000 &  0.000 &  0.000 &  0.000 &  0.000 &  0.000 &  0.000 &  0.000\\ 
        0.575 & 2.566 &  0.063 &  0.090 &  0.036 &  0.025 &  0.055 &  0.065 &  0.015 &  0.005\\ 
        0.600 & 2.609 &  0.130 &  0.177 &  0.076 &  0.053 &  0.118 &  0.129 &  0.034 &  0.014\\ 
        0.625 & 2.655 &  0.202 &  0.268 &  0.123 &  0.086 &  0.191 &  0.198 &  0.061 &  0.026 \\
        0.650 & 2.693 &  0.269 &  0.361 &  0.178 &  0.117 &  0.263 &  0.270 &  0.095 &  0.036 \\
      \hline
    \end{tabular}
  \end{center}
\end{table*}

\begin{table*}
  \begin{center}
    \caption{K and K+e corrections for SDSS LRGs to z = 0.55}
    \begin{tabular}{ c  c  c  c  c  c  c  c  c  c  c  c  c  c  c} 
        \hline  
      \multicolumn{1}{c}{} &
      \multicolumn{7}{c|}{Passive} &
      \multicolumn{7}{c}{Star-forming} \\
      \multicolumn{1}{c}{} &
      \multicolumn{1}{c}{} &
      \multicolumn{3}{c}{K} &
      \multicolumn{3}{c}{K+e} &
      \multicolumn{1}{c}{} &
      \multicolumn{3}{c|}{K} &
      \multicolumn{3}{c}{K+e} \\
      \multicolumn{1}{c}{z} &
      \multicolumn{1}{c}{$g-i$} &
      \multicolumn{1}{c}{$u\rightarrow g$}&
      \multicolumn{1}{c}{$g\rightarrow r$}&
      \multicolumn{1}{c}{$r\rightarrow i$}&
      \multicolumn{1}{c}{$u\rightarrow g$}&
      \multicolumn{1}{c}{$g\rightarrow r$}&
      \multicolumn{1}{c}{$r\rightarrow i$}&
      \multicolumn{1}{c}{$g-i$} &
      \multicolumn{1}{c}{$u\rightarrow g$}&
      \multicolumn{1}{c}{$g\rightarrow r$}&
      \multicolumn{1}{c}{$r\rightarrow i$}&
      \multicolumn{1}{c}{$u\rightarrow g$}&
      \multicolumn{1}{c}{$g\rightarrow r$}&
      \multicolumn{1}{c}{$r\rightarrow i$}\\
      \hline 
        0.150 & 1.697 &  0.200 &  0.031 & -0.012 &  0.727 &  0.501 &  0.386 & 1.597 &  0.381 &  0.067 & -0.004 &  0.850 &  0.520 &  0.387\\
        0.175 & 1.813 &  0.391 &  0.170 &  0.031 &  0.877 &  0.607 &  0.401 & 1.703 &  0.491 &  0.188 &  0.035 &  0.944 &  0.617 &  0.402\\
        0.200 & 1.929 &  0.598 &  0.311 &  0.074 &  1.039 &  0.715 &  0.416 & 1.808 &  0.598 &  0.311 &  0.074 &  1.039 &  0.715 &  0.416\\
        0.225 & 2.034 &  0.800 &  0.443 &  0.116 &  1.198 &  0.813 &  0.431 & 1.903 &  0.703 &  0.434 &  0.118 &  1.113 &  0.802 &  0.429\\
        0.250 & 2.121 &  1.020 &  0.565 &  0.162 &  1.375 &  0.900 &  0.448 & 1.982 &  0.830 &  0.573 &  0.176 &  1.159 &  0.876 &  0.441\\
        0.275 & 2.208 &  1.245 &  0.687 &  0.214 &  1.562 &  0.987 &  0.472 & 2.060 &  0.920 &  0.685 &  0.227 &  1.237 &  0.954 &  0.463\\
        0.300 & 2.300 &  1.477 &  0.817 &  0.264 &  1.765 &  1.082 &  0.495 & 2.143 &  0.992 &  0.793 &  0.271 &  1.319 &  1.040 &  0.484\\
        0.325 & 2.393 &  1.714 &  0.955 &  0.316 &  1.973 &  1.181 &  0.520 & 2.225 &  1.054 &  0.906 &  0.318 &  1.393 &  1.128 &  0.507\\
        0.350 & 2.450 &  1.931 &  1.064 &  0.377 &  2.164 &  1.252 &  0.553 & 2.275 &  1.096 &  0.995 &  0.372 &  1.448 &  1.192 &  0.538\\ 
      \hline
    \end{tabular}
  \end{center}
\end{table*}

\begin{table*}
  \begin{center}
    \caption{K and K+e corrections for 2SLAQ LRGs to z = 0.2}
    \begin{tabular}{ c  c  c  c  c  c  c  c  c  c  c  c  c  c  c} 
        \hline  
      \multicolumn{1}{c}{} &
      \multicolumn{7}{c|}{Passive} &
      \multicolumn{7}{c}{Star-forming} \\
      \multicolumn{1}{c}{} &
      \multicolumn{1}{c}{} &
      \multicolumn{3}{c}{K} &
      \multicolumn{3}{c}{K+e} &
      \multicolumn{1}{c}{} &
      \multicolumn{3}{c|}{K} &
      \multicolumn{3}{c}{K+e} \\
      \multicolumn{1}{c}{z} &
      \multicolumn{1}{c}{$g-i$} &
      \multicolumn{1}{c}{$r\rightarrow g$}&
      \multicolumn{1}{c}{$i\rightarrow r$}&
      \multicolumn{1}{c}{$z\rightarrow i$}&
      \multicolumn{1}{c}{$r\rightarrow g$}&
      \multicolumn{1}{c}{$i\rightarrow r$}&
      \multicolumn{1}{c}{$z\rightarrow i$}&
      \multicolumn{1}{c}{$g-i$} &
      \multicolumn{1}{c}{$r\rightarrow g$}&
      \multicolumn{1}{c}{$i\rightarrow r$}&
      \multicolumn{1}{c}{$z\rightarrow i$}&
      \multicolumn{1}{c}{$r\rightarrow g$}&
      \multicolumn{1}{c}{$i\rightarrow r$}&
      \multicolumn{1}{c}{$z\rightarrow i$}\\
      \hline 
        0.450 & 2.574 & -0.781 & -0.214 & -0.183 & -1.043 & -0.458 & -0.408 & 2.376 & -0.776 & -0.227 & -0.192 & -1.016 & -0.449 & -0.402\\ 
        0.475 & 2.635 & -0.662 & -0.185 & -0.163 & -0.959 & -0.453 & -0.409 & 2.422 & -0.672 & -0.201 & -0.173 & -0.938 & -0.445 & -0.403\\ 
        0.500 & 2.682 & -0.538 & -0.149 & -0.146 & -0.872 & -0.441 & -0.414 & 2.455 & -0.563 & -0.168 & -0.157 & -0.856 & -0.434 & -0.407\\ 
        0.525 & 2.735 & -0.421 & -0.115 & -0.124 & -0.789 & -0.432 & -0.413 & 2.491 & -0.460 & -0.138 & -0.137 & -0.778 & -0.425 & -0.406\\ 
        0.550 & 2.788 & -0.311 & -0.074 & -0.097 & -0.715 & -0.416 & -0.408 & 2.526 & -0.364 & -0.101 & -0.112 & -0.708 & -0.410 & -0.401\\ 
        0.575 & 2.850 & -0.208 & -0.034 & -0.070 & -0.647 & -0.400 & -0.403 & 2.566 & -0.274 & -0.065 & -0.087 & -0.643 & -0.395 & -0.396\\ 
        0.600 & 2.921 & -0.108 &  0.010 & -0.039 & -0.579 & -0.380 & -0.394 & 2.609 & -0.187 & -0.025 & -0.059 & -0.579 & -0.376 & -0.387\\ 
        0.625 & 3.001 & -0.005 &  0.063 & -0.004 & -0.507 & -0.352 & -0.381 & 2.655 & -0.096 &  0.022 & -0.026 & -0.510 & -0.349 & -0.375 \\
        0.650 & 3.077 &  0.103 &  0.126 &  0.029 & -0.431 & -0.315 & -0.371 & 2.693 & -0.003 &  0.077 &  0.005 & -0.438 & -0.315 & -0.365 \\
      \hline
    \end{tabular}
  \end{center}
\end{table*}

\label{lastpage}

\end{document}